\documentclass[aps,prc,preprint,groupedaddress,showpacs]{revtex4}
\usepackage[dvips]{graphicx}
\usepackage{amsmath}

\begin{document}
\title{Structure of Strange Dwarfs with Color Superconducting Core}

\author{Masayuki Matsuzaki}
\email[]{matsuza@fukuoka-edu.ac.jp}
\affiliation{Department of Physics, Fukuoka University of Education, 
             Munakata, Fukuoka 811-4192, Japan}
\author{Etsuchika Kobayashi}
\affiliation{Department of Physics, Fukuoka University of Education, 
             Munakata, Fukuoka 811-4192, Japan}


\begin{abstract}
 We study effects of two-flavor color superconductivity on the structure of 
strange dwarfs, which are stellar objects with similar masses and radii with 
ordinary white dwarfs but stabilized by the strange quark matter core. We find 
that unpaired quark matter is a good approximation to the core of strange dwarfs. 
\end{abstract}

\maketitle

 Witten made a conjecture that the absolute ground state of quantum chromodynamics 
(QCD) is not $^{56}$Fe but strange quark matter, which is a plasma composed of 
almost equal number of deconfined u, d, and s quarks~\cite{Wit}. Although this 
conjecture has been neither confirmed nor rejected, if this is true, since 
deconfinement is expected in high density cores of compact stars, there could 
exist stars that contain strange quark matter converted from two-flavor quark 
matter via weak interaction. Strange quark stars whose radii are about 10 km, 
with or without {\it thin} nuclear crust, have long been investigated. 

 Glendenning et al. proposed a new class of compact stars containing strange 
quark matter and {\it thick} nuclear crust ranging from a few hundred to ten 
thousand km~\cite{Gle1,Gle2,Gle3}. They named them the strange dwarfs because 
their radii correspond to those of white dwarfs. Alcock et al. discussed the 
mechanism that the strange quark core supports the cruct~\cite{Alc}. 
Since the mass of s quark is larger than those of u and d, strange quark matter 
is positively charged. In order to electrically neutralize the core, electrons 
are bound to the surface of the core. They estimated that the thickness of this 
electric dipole layer is a few hundred fm. Then this layer can support a nuclear 
crust. Although Alcock et al. considered only thin crusts, Glendenning et al. 
considered thick crusts up to about ten thousand km. 
Very recently, Mathews et al. identified eight candidates of strange dwarfs 
from observed data~\cite{Mat}.

 A theoretical facet whose importance in nuclear physics was recognized later 
is color superconductivity in quark matter. 
At asymptotically high density, the color-flavor locking (CFL) 
is believed to be the ground state~\cite{Raj}. At realistic densities, however, 
the two-flavor color superconductivity (2SC) is thought to be realized even when 
electric neutrality is imposed if the coupling constant is strong~\cite{Abu}.
Thus, in the present paper, we discuss effects of the 2SC phase in the strange 
quark matter core on the structure of strange dwarfs. 

 In order to determine the structure of compact stars, we solve the general 
relativistic Tolman-Oppenheimer-Volkoff (TOV) equation, 
\begin{gather}
\frac{dp(r)}{dr}=-\frac{G\epsilon(r)M(r)}{c^2r^2}
\left(1+\frac{p(r)}{\epsilon(r)}\right)
\left(1+\frac{4\pi r^3p(r)}{M(r)c^2}\right)
\left(1-\frac{2GM(r)}{c^2r}\right)^{-1} , \\
M(r)=4\pi\int_0^r\frac{\epsilon(r')}{c^2}r'^2dr' ,
\end{gather}
for the pressure $p(r)$, the energy density $\epsilon(r)$, and the mass enclosed 
within the radius $r$, $M(r)$. Here $G$ is the gravitational constant and $c$ is 
the speed of light. The equation is closed when an equation of state (EOS), a 
relation between $p$ and $\epsilon$, is specified. In the present case, strange 
dwarfs are composed of the strange quark matter core and the nuclear crust. 
Accordingly two parameters, the pressure at the center and at the core-crust 
boundary, must be specified to integrate the TOV equation. The latter must be 
equal or less than that corresponds to the nucleon drip density 
$\epsilon_\mathrm{drip}$. Otherwise neutrons drip and gravitate to the core. 
In the present calculation we take a $p_\mathrm{cruct}$ calculated from 
$\epsilon_\mathrm{crust}=\epsilon_\mathrm{drip}$. 

 We assume zero temperature throughout this paper. As for the EOS of the quark 
core, we adopt the MIT bag model without any QCD corrections (see 
Ref.~\cite{Gle3}, for example). For unpaired free quark matter, 
\begin{gather}
p=-B+\sum_f\frac{1}{4\pi^2}
\left[\mu_fk_{\mathrm{F}f}\left(\mu_f^2-\frac{5}{2}m_f^2\right)
      +\frac{3}{2}m_f^4\ln\left(\frac{\mu_f+k_{\mathrm{F}f}}{m_f}\right)\right] , \\
\epsilon=B+\sum_f\frac{3}{4\pi^2}
\left[\mu_fk_{\mathrm{F}f}\left(\mu_f^2-\frac{1}{2}m_f^2\right)
      -\frac{1}{2}m_f^4\ln\left(\frac{\mu_f+k_{\mathrm{F}f}}{m_f}\right)\right] , 
\end{gather}
where $m_f$, $k_{\mathrm{F}f}$, and $\mu_f=\sqrt{m_f^2+k_{\mathrm{F}f}^2}$ are 
the mass, the Fermi momentum, and the chemical potential of quarks of each flavor, 
respectively, and $f$ runs $u$, $d$, and $s$. Hereafter we put $c=\hbar=1$. 
The quantity $B$ is the bag constant. 
The effect of color superconductivity is incorporated as a chemical potential 
dependent effective bag constant. In the 2SC case~\cite{Alf}, 
\begin{equation}
B_\mathrm{eff}=B-\frac{1}{\pi^2}\Delta^2(\mu)\mu^2 ,
\end{equation}
where $\Delta(\mu)$ is the quark pairing gap as a function of a chemical potential 
$\mu$, whose relation to $\mu_f$ is specified later. 

 The pairing gap is obtained as a function of the Fermi momentum by solving the 
gap equation~\cite{MM}
\begin{gather}
\Delta(k_\mathrm{F})
=-\frac{1}{8\pi^2}  \int_0^\infty \bar v(k_\mathrm{F},k)
  \frac{\Delta(k)}{E'(k)}k^2dk , \\
E'(k)=\sqrt{(E_k-E_{k_\mathrm{F}})^2+3\Delta^2(k)} ,
\end{gather}
with $k_\mathrm{F}=k_{\mathrm{F}u}=k_{\mathrm{F}d}$, $E_k=\sqrt{k^2+m_q^2}$, and 
$m_q=m_u=m_d$. The one gluon exchange pairing interaction is given by 
\begin{gather}
\bar v(p,k)=-\frac{\pi}{3}\alpha_\mathrm{s}\frac{1}{pkE_pE_k}
\bigg(\left(2E_pE_k+2m_q^2+p^2+k^2+m_\mathrm{E}^2\right)
       \ln\left(\frac{(p+k)^2+m_\mathrm{E}^2}{(p-k)^2+m_\mathrm{E}^2}\right)
\biggr. \nonumber \\
\biggl.
+2\left(6E_pE_k-6m_q^2-p^2-k^2\right)
       \ln \bigg|\frac{p+k}{p-k} \bigg|
\bigg) , \nonumber \\
m_\mathrm{E}^2=\frac{4}{\pi}\alpha_\mathrm{s}\mu^2 ,
\end{gather}
where $p$ and $k$ are the magnitudes of 3-momenta. 
The running coupling constant is given by~\cite{Hig} 
\begin{gather}
\alpha_\mathrm{s}({\bf q}^2)=\frac{4\pi}{9}
   \frac{1}{\ln\left(\frac{q_\mathrm{max}^2+q_\mathrm{c}^2}
                          {\Lambda_\mathrm{QCD}^2}
               \right)} , \nonumber \\
{\bf q}={\bf p}-{\bf k} , \nonumber \\
 q_\mathrm{max}=\max\{p,k\} .
\end{gather}
As for the EOS of the crust, we adopt the tabulated one for 
$\beta$-equilibrium nuclear matter of Baym, Pethick, and Sutherland~\cite{Bay} 
(BPS) conforming to Refs.~\cite{Gle1,Gle2,Gle3}. 

 The positively charged strange quark matter in the core is simply approximated 
by $\mu=\mu_u=\mu_d=\mu_s$. Quark masses are given by $m_u=m_d=$ 10 MeV, 
$m_s=$ 150 MeV. The bag constant is chosen to be $B^{1/4}=$ 160 MeV. Parameters 
entering into the pairing interaction are $q_\mathrm{c}^2=1.5\Lambda_\mathrm{QCD}^2$ 
and $\Lambda_\mathrm{QCD}=$ 400 MeV. The nucleon drip density is 
$\epsilon_\mathrm{drip}=$ 4.3$\times$10$^{11}$ g/cm$^3$. 

\begin{figure}[htbp]
\begin{center}
\includegraphics[width=7cm]{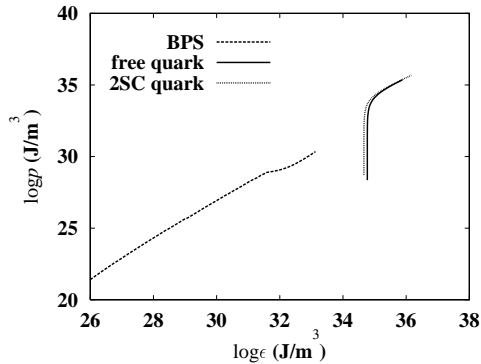}
\end{center}
\caption{Equations of state of free and 2SC quark matter and 
$\beta$-equilibrium nuclear matter. The latter is tabulated in Refs.~\cite{Bay} 
and \cite{Gle3}.}
\label{fig1}
\end{figure}%

\begin{figure}[htbp]
\includegraphics[width=7cm]{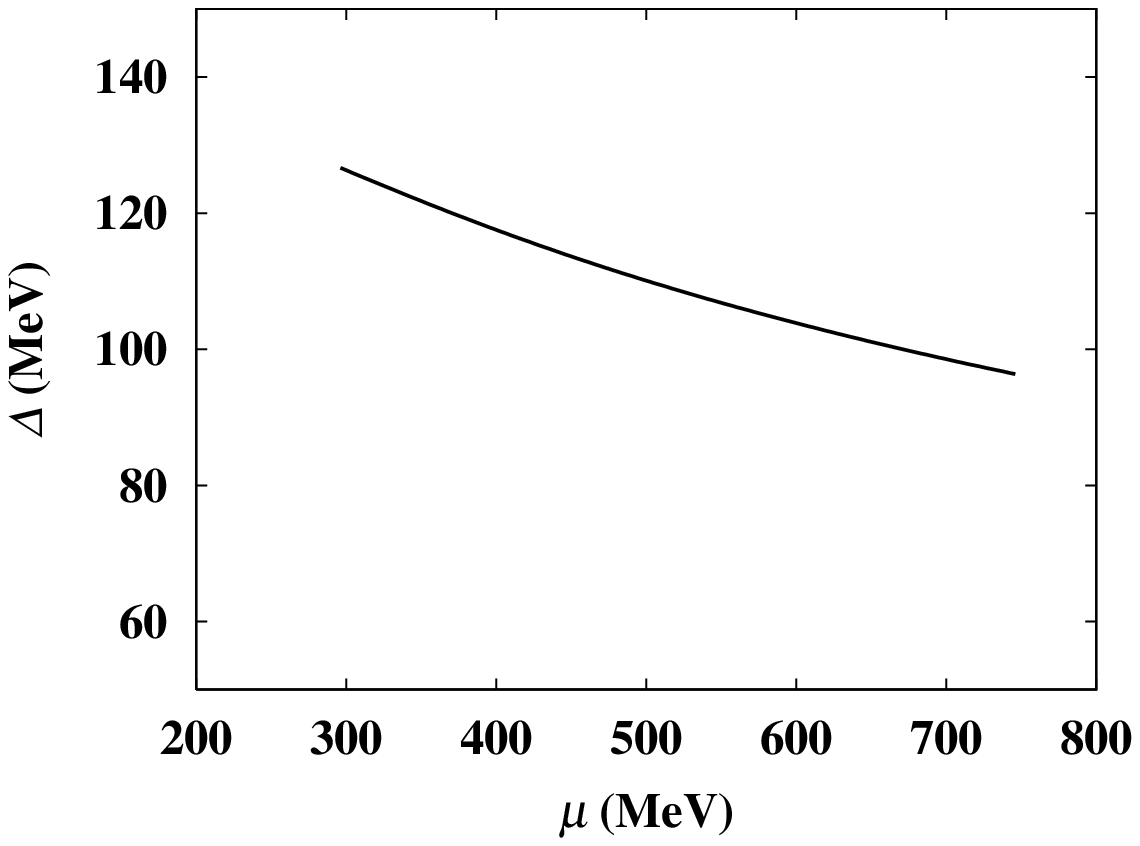}
\includegraphics[width=7cm]{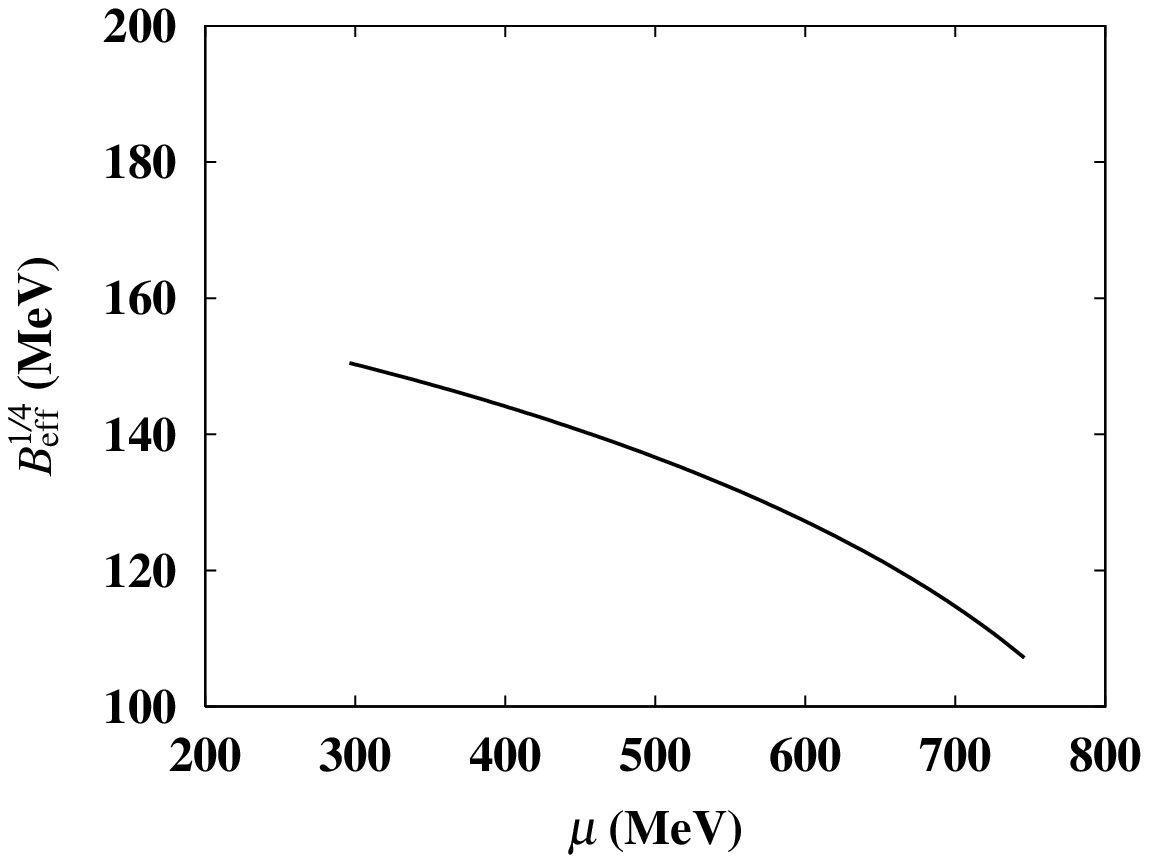}
\caption{Left: color superconducting pairing gap and right: effective bag constant, 
as functions of the quark chemical potential.}
\label{fig2}
\end{figure}

 The adopted EOS is displayed in Fig.~\ref{fig1}. The logarithm is to base 10 
throughout this paper. The quark matter EOS describes the core and the BPS EOS 
describes the crust. At the boundary, the pressure is common whereas the energy 
density jumps discontinuously. In order to obtain the EOS for 2SC matter, 
the pairing gap must be calculated at each $k_\mathrm{F}$ beforehand. This is 
shown in Fig.~\ref{fig2} left. The effective bag constant determined by the 
pairing gap is shown in Fig.~\ref{fig2} right. The resulting 2SC EOS is included 
in Fig.~\ref{fig1}. 

\begin{figure}[htbp]
\begin{center}
\includegraphics[width=7cm]{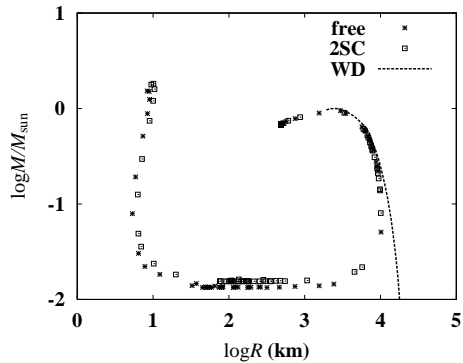}
\end{center}
\caption{Mass-radius relation of strange dwarfs and white dwarfs.}
\label{fig3}
\end{figure}%

\begin{figure}[htbp]
\includegraphics[width=7cm]{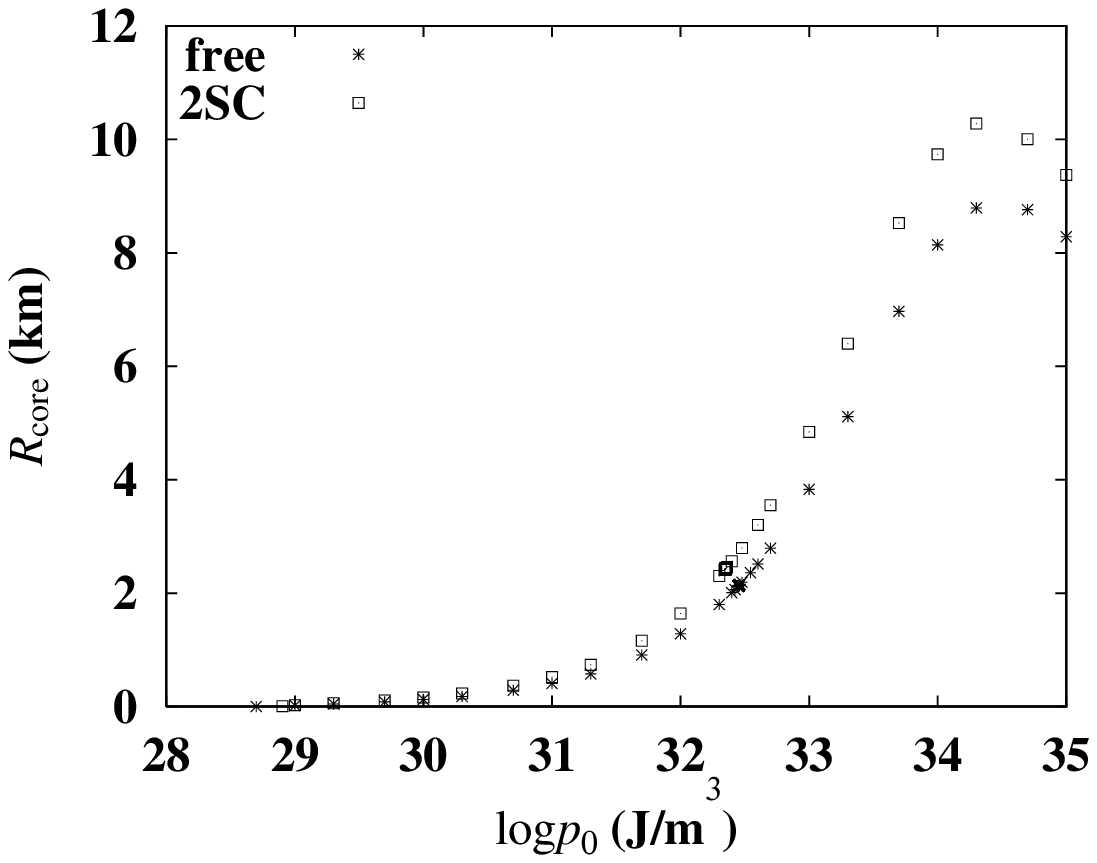}
\includegraphics[width=7cm]{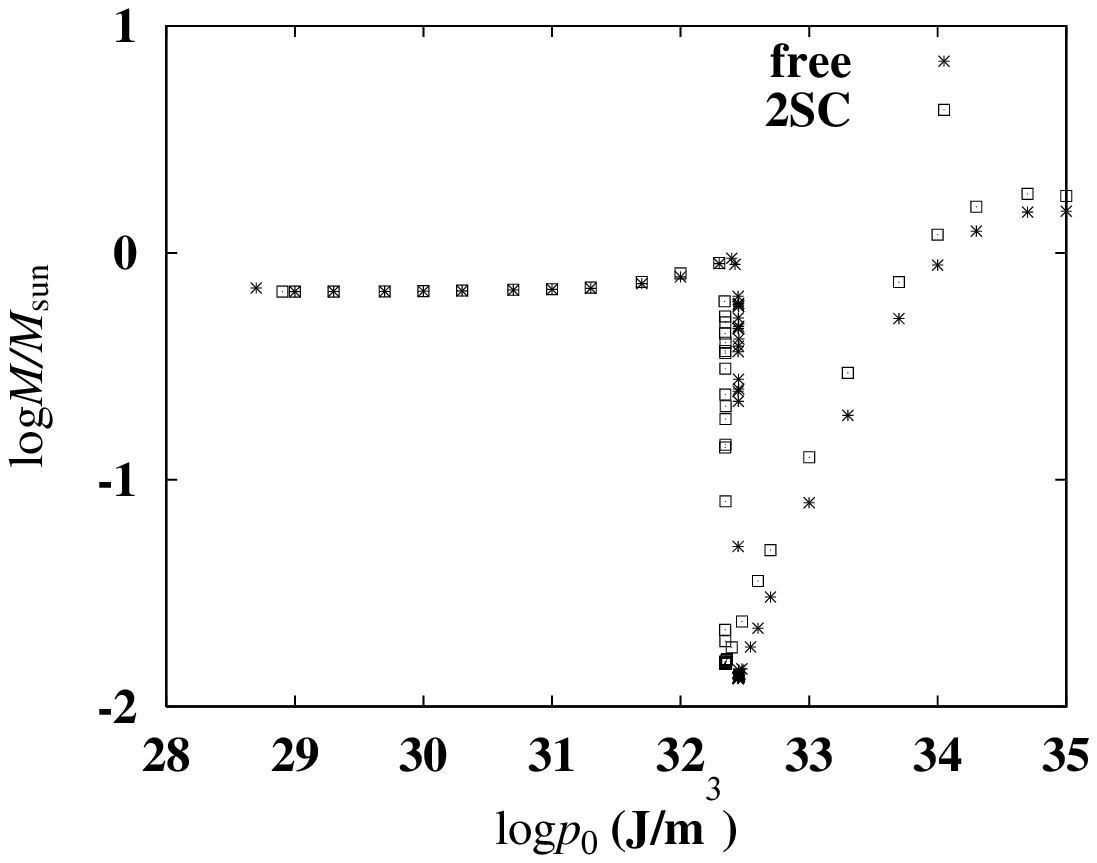}
\caption{Left: core radius and right: mass of strange dwarfs, as functions of the 
central pressure.}
\label{fig4}
\end{figure}

\begin{figure}[htbp]
\includegraphics[width=7cm]{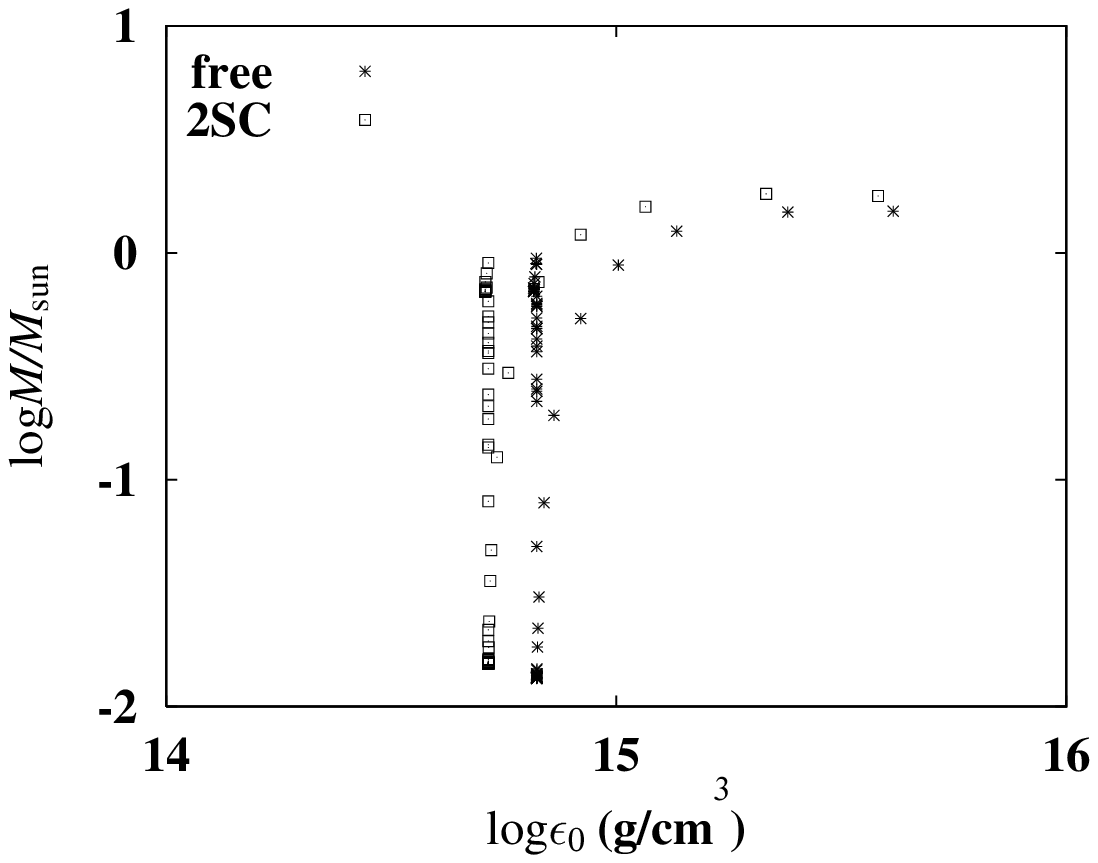}
\includegraphics[width=7cm]{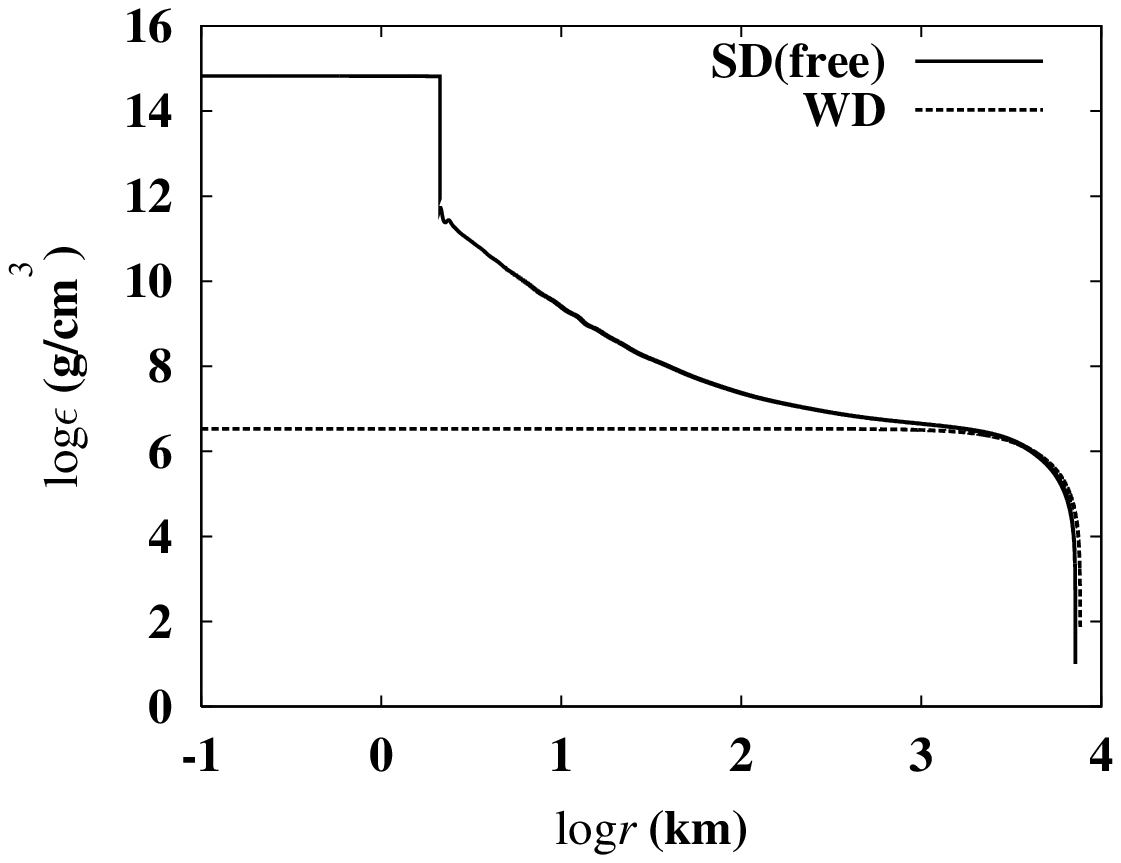}
\caption{Left: mass of strange dwarfs as a function of the central energy density. 
Right: energy profile of a strange dwarf with $M/M_\mathrm{sun}=$ 0.465 and that of 
a white dwarf with $M/M_\mathrm{sun}=$ 0.466.}
\label{fig5}
\end{figure}

 Figure~\ref{fig3} presents the mass-radius relation obtained by integrating the 
TOV equation with a fixed $p_\mathrm{cruct}$, determined from 
$\epsilon_\mathrm{crust}=\epsilon_\mathrm{drip}$, and various central pressures. 
This result can be classified into three regions. The first region (larger 
central pressures), almost vertical curve at around $R\sim$ 10 km, describes strange 
stars with thin crusts. In this region, color superconductivity makes the maximum 
mass and radius larger because the pairing gap reduces the bag constant and 
consequently the energy density decreases and the pressure increases. 
This is consistent with another calculation with the CFL 
phase~\cite{Lug}. The second region, horizontal at around 
$M/M_\mathrm{sun}\sim$ 10$^{-2}$, and the third region, vertical at around 
$R\sim$ 10$^4$ km up to the maximum mass, correspond to strange dwarfs. In the second 
region, color superconducting quark cores support slightly larger masses than 
unpaired free quark cores. In the third region, effect of color superconductivity 
is negligible. In Fig.~\ref{fig3}, The mass-radius relation of ordinary white 
dwarfs without quark matter cores calculated by adopting the BPS EOS is also shown 
although it is known that the BPS EOS is not very suitable for white dwarfs. 
As the central pressure decreases, the quark matter core shrinks 
(Fig.~\ref{fig4} left) and eventually strange dwarfs reduce to ordinary white dwarfs. 
When their masses are the same, the former is more compact than the latter 
(see also Fig.~\ref{fig5} right) because of the gravity of the core. Mathews et al. 
paid attention to this difference in the mass-radius relation and 
classified the observed data of dwarfs~\cite{Mat}. According to their work, eight of 
them are classified into strange dwarfs.

 Figure~\ref{fig4} right indicates that strange dwarfs, in particular those of 
10$^3$ km $< R <$ 10$^4$ km, are realized in a very narrow range of the central 
pressure. This is reflected in the density of calculated points. During this 
rapid structure change from the second to the third region, the core radius 
almost does not change, see Fig.~\ref{fig4} left. Figure~\ref{fig5} left also 
graphs $M/M_\mathrm{sun}$ as Fig.~\ref{fig4} right but as a function of the 
central energy density. The difference between these two figures at the low 
pressure/energy density side can be understood from the quark matter EOS in 
Fig.~\ref{fig1} such that the pressure decreases steeply at the lowest energy 
density. Figure~\ref{fig5} left indicates that strange dwarfs have central 
energy densities just below the lowest stable compact strange stars and several 
orders of magnitude larger than those of ordinary white dwarfs. This is clearly 
demonstrated in Fig.~\ref{fig5} right. 

 To summarize, we have solved the Tolman-Oppenheimer-Volkoff equation for strange 
dwarfs with $\epsilon_\mathrm{crust}=\epsilon_\mathrm{drip}$ and a wide range of 
the central pressure. We have examined effects of the two-flavor color 
superconductivity in the strange quark matter core in a simplified manner. 
The obtained results indicate that, aside from a slight increase of the minimum 
mass, effect of color superconductivity is negligible in the mass-radius 
relation. This is consistent with the conjecture given in Ref.~\cite{Mat}. 
As a function of the central energy density, however, strange dwarfs are realized 
at slightly lower energy densities than the unpaired free quark case reflecting 
the effect on the equation of state. Recently Usov discussed that electric 
fields are also generated on the surface of the color-flavor locked 
matter~\cite{Uso}. This suggests that strange dwarfs with color-flavor locked 
cores might also be possible although this is expected only at relatively high 
densities. Since the pairing gap enters into the calculation only through the 
effective bag constant, aside from a possible slight change in chemical 
potentials, it can surely be expected that the effect of color-flavor locking 
does not differ much from that of the two-flavor color superconductivity. 
In conclusion, unpaired quark matter is a good approximation to the core 
of strange dwarfs. Another aspect that might be affected by color 
superconductivity is the cooling~\cite{Ben}. This is beyond the scope of the 
present study.

\end{document}